\documentclass[sort&compress]{aip-cp}

\usepackage[numbers]{natbib}
\usepackage{rotating}
\usepackage{graphicx}
\usepackage{color}

\usepackage{amssymb}
\usepackage{subeqn}
\usepackage{bm}

\def\bal#1\eal{\begin{align}#1\end{align}}
\newcommand\beq{\begin{equation}}
\newcommand\eeq{\end{equation}}
\newcommand\beqa{\begin{eqnarray}}
\newcommand\eeqa{\end{eqnarray}}

\newcommand{\text}[1]{\textrm{\scriptsize{#1}}}
\newcommand{\eqref}[1]{(\ref{#1})}

\newcommand{\thr}{\text{th}}

\begin{document}

\title{Kinetic Theory of Soft Matter. The Penetrable-Square-Well Model}

\author[aff1]{Jos\'e Luis S\'anchez-Tena
}
\author[aff1,aff2]{Andr\'es Santos\corref{cor1}
}
\eaddress[url]{http://www.eweb.unex.es/eweb/fisteor/andres/Cvitae/}
\author[aff3]{Pablo Pajuelo
}

\affil[aff1]{Departamento de F\'isica, Universidad de Extremadura, 06006 Badajoz, Spain.}
\affil[aff2]{Instituto de Computaci\'on Cient\'ifica Avanzada (ICCAEx), Universidad de Extremadura, 06006 Badajoz, Spain.}
\affil[aff3]{Android Development Team, QUADRAM, 28045 Madrid, Spain}
\corresp[cor1]{Corresponding author: andres@unex.es}

\maketitle

\begin{abstract}
The penetrable-square-well (PSW) pair interaction potential is defined as $\phi (r)=\epsilon_r$ if the two interacting particles are overlapped ($r<\sigma$),  $\phi(r)=-\epsilon_a$ inside a  corona ($\sigma <r<\lambda$), and $\phi(r)=0$ otherwise ($r>\lambda$). Thus, the potential reduces to the conventional square-well (SW)  one in the limit  $\epsilon_r\to\infty$ and to the penetrable-sphere (PS) potential if  $\epsilon_a\to0$ or  $\lambda\to\sigma$.
This paper aims at studying the temperature dependence of the Navier--Stokes transport coefficients of a dilute gas of particles interacting via the PSW model. By exploiting the fact that the PSW scattering process  is analogous to that of a light ray passing through two concentric spherical media with different refractive indices, the scattering angle is analytically derived as a function of the impact parameter and the relative velocity of the colliding particles; depending on the values of those two quantities, collisions can be soft, hard, or grazing. Next, by standard application of known general results from the Chapman--Enskog method, the Navier--Stokes transport coefficients in the first-order approximation are numerically evaluated. It is found that the PSW coefficients are practically indistinguishable from the SW ones for temperatures low enough ($k_BT\lesssim 0.2 \epsilon_r$), there exists a transition regime ($0.2 \epsilon_r\lesssim k_BT\lesssim 10\epsilon_r$) where the transport coefficients interpolate between the SW and the PS ones, and finally the PSW coefficients are comparable to the PS ones for high enough temperatures ($k_BT\gtrsim 10\epsilon_r$). The results are applied to the temperature profiles of the planar Fourier flow.

\end{abstract}

\section{INTRODUCTION}

As is well known, the standard kinetic theory of gases is usually applied to particles interacting via unbounded spherically symmetric pair potentials, such as hard spheres, power-law repulsive interactions, the square-well (SW) model, or the Lennard-Jones potential \cite{CC70,FK72,C88,GS03}. On the other hand, the equilibrium properties of so-called ``soft-matter'' fluids of particles interacting with bounded pair potentials are the subject of an increasing interest as models of colloidal systems, such as micelles in a solvent or star copolymer suspensions \cite{C62,MW89,SS97,LWL98,LLWL01,L01,SF02,AS04,MYS07,PSMG14,MP15,MPP18}. Nonetheless, the somewhat smaller attention paid to the  nonequilibrium transport properties of those systems has been practically restricted to purely repulsive interactions \cite{A75,GDL79,S05b,MM06,MM07,MM09,AMS09,KKMET09,PKSET09,SE10,SKKS10}.

\begin{figure}[ht]
\centerline{\includegraphics[scale=0.4]{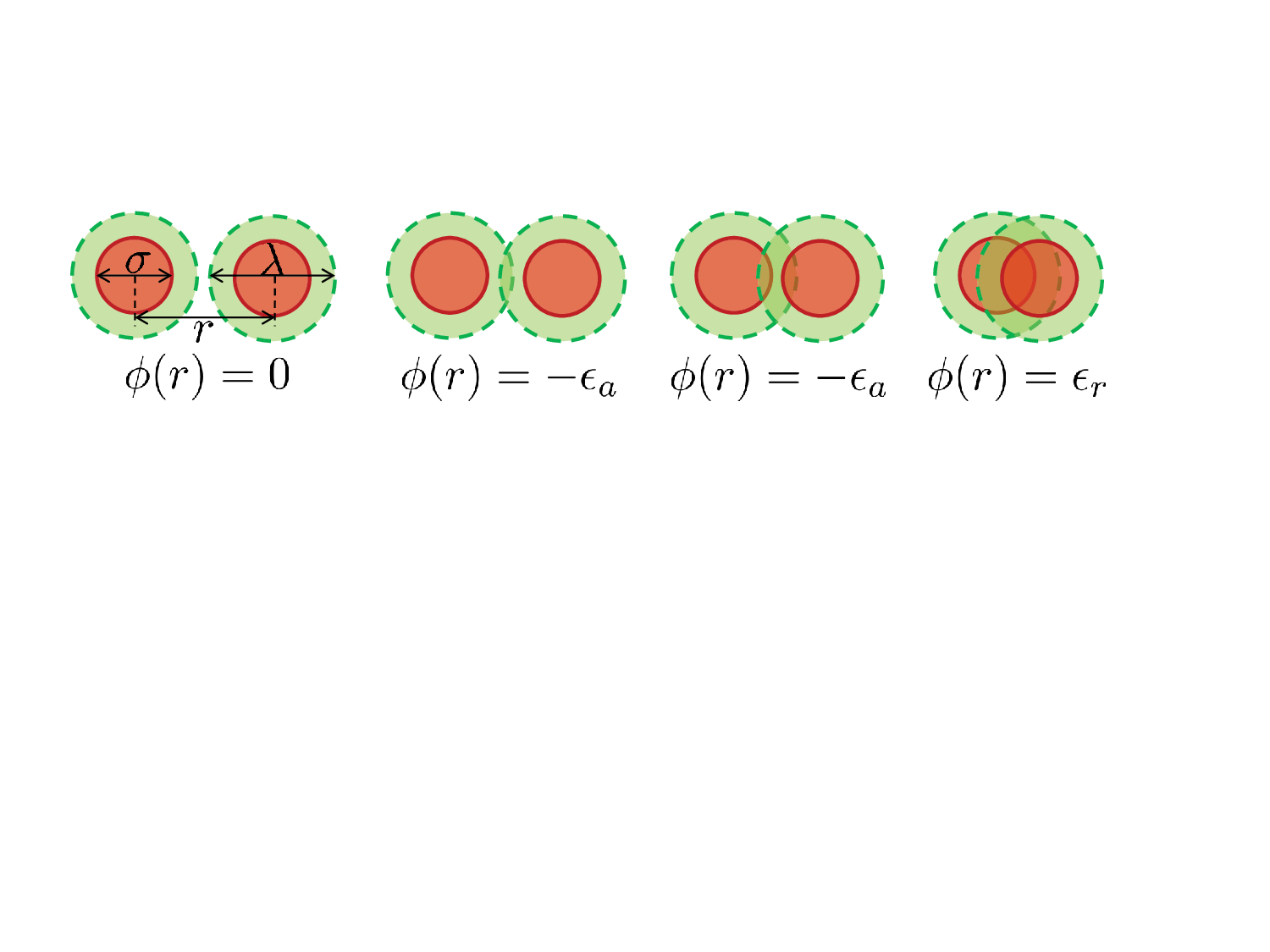}}
\caption{Sketch of the PSW interaction potential for four representative configurations.}
\label{sketch}
\end{figure}

The aim of this paper is to contribute to the understanding of the nonequilibrium properties of a dilute ``gas'' made of particles interacting via bounded potentials by considering the simplest model that combines a soft repulsive part and an attractive tail, namely the
penetrable-square-well (PSW) model \cite{SFG08,FGMS09,FGMS10,FMSG11a,FMSG11b,M14}.
The PSW interaction potential is defined as
\beq
\label{PSW}
\phi(r)=\left\{
\begin{array}{ll}
  \epsilon_r,&r<\sigma,\\
  -\epsilon_a,&\sigma<r<\lambda,\\
  0,&r>\lambda.
\end{array}
\right.
\eeq
This potential presents a \emph{repulsive} soft core with a finite barrier $\epsilon_r$, thus allowing for overlapping configurations ($r<\sigma$), plus an \emph{attractive} well of depth $-\epsilon_a$ inside the corona ($\sigma<r<\lambda$).  A sketch is shown in Fig.~\ref{sketch}. Therefore, in the PSW model the gas behaves as a SW gas \cite{HH51,LHV58,HCB64,KS83,KSK89} in the limit  $\epsilon_r\to\infty$   and as penetrable spheres (PS) \cite{KS83,S05b,SKKS10} if  $\epsilon_a\to0$ or  $\lambda\to\sigma$.

The scattering angle as a function of the relative velocity and of the impact parameter is derived in this paper by taking into account that the scattering process associated with the PSW potential is analogous to that of a light ray passing through two concentric spherical media with different refractive indices. Next, by a standard application of the Chapman--Enskog method \cite{CC70,FK72}, the Navier--Stokes shear viscosity, thermal conductivity, and self-diffusion coefficients of the PSW dilute gas are  numerically evaluated in the first-order Sonine approximation.

\section{SCATTERING PROCESS}
In a two-body collision,  we can consider the equivalent one-body problem in which a projectile particle (with a reduced mass $\mu=m/2$, $m$ being the mass of each colliding particle) feels a central potential $\phi(r)$ centered at the origin. The projectile  approaches the ``target'' with a (relative) speed $g$ and an impact parameter $b$, being  deflected after interaction with a scattering angle $\chi(b,g)$.

In the case of the PSW potential, it is obvious that the impact parameter, $b$, must be smaller than the diameter of the corona, $\lambda$, for a true collision to take place. The incoming kinetic energy of the reduced mass is $\frac{1}{2}\mu g^2$ but, once the projectile enters into the corona, its kinetic energy changes to  $\frac{1}{2}\mu g_a^2=\frac{1}{2}\mu g^2+\epsilon_a$.
In case the projectile penetrates the inner repulsive core, its kinetic energy changes to $\frac{1}{2}\mu g_r^2=\frac{1}{2}\mu g^2-\epsilon_r=\frac{1}{2}\mu g_a^2-\epsilon_a-\epsilon_r$. Therefore, relative to the incoming speed, the speed in the corona increases by  a  factor $n_a(g)\equiv g_a/g$, while the speed inside the core decreases by  a factor $n_r(g)\equiv g_r/g$, where
\beq
n_a(g)=\sqrt{1+{4\epsilon_a}/{m g^2}},\quad n_r(g)=\sqrt{1-{4\epsilon_r}/{m g^2}}.
\eeq
For further use, let us also introduce the threshold values
\beq
g_a^\thr=2\sqrt{\frac{\epsilon_a/m}{\lambda^2/\sigma^2-1}},\quad g_r^\thr=2\sqrt{\epsilon_r/m}.
\eeq
Note that $n_a(g_a^\thr)=\lambda/\sigma$ and $n_r(g_r^\thr)=0$. If $\epsilon_r/\epsilon_a$ and/or $\lambda/ \sigma$ are sufficiently large so that $\epsilon_r/\epsilon_a>\left(\lambda^2/\sigma^2-1\right)^{-1}$, then $g_a^\thr<g_r^\thr$; otherwise, $g_a^\thr>g_r^\thr$.

\begin{figure}[ht]
\begin{tabular}{cc}
 \includegraphics[scale=0.7]{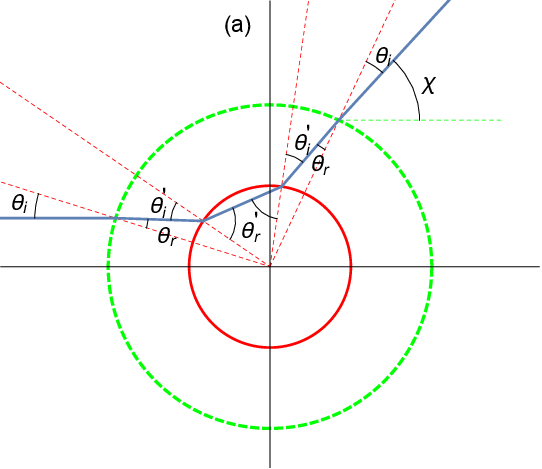}&\includegraphics[scale=0.7]{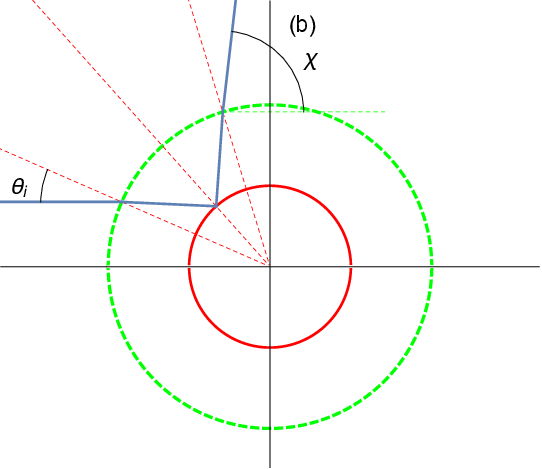}\\
\includegraphics[scale=0.7]{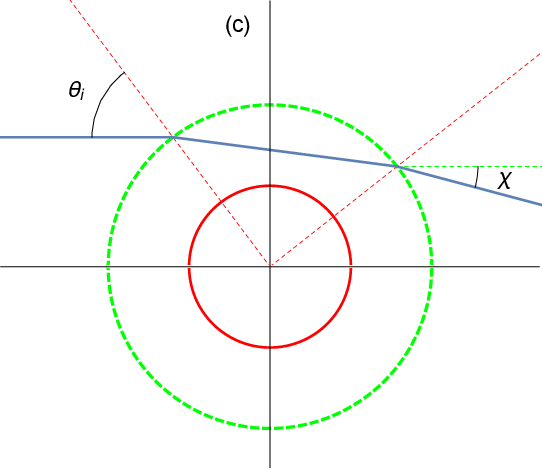}&\includegraphics[scale=0.7]{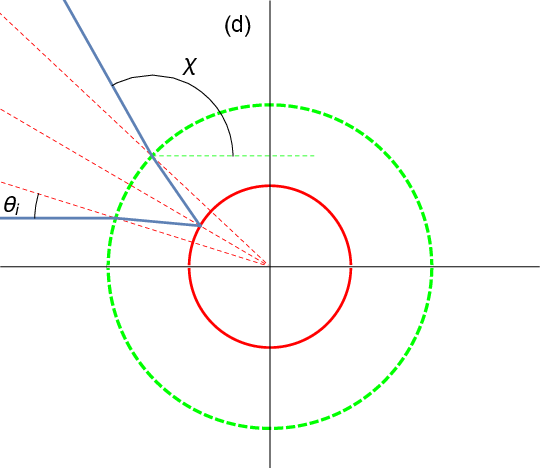}
\end{tabular}
\caption{Representative scattering processes for a PSW potential with $\epsilon_r/\epsilon_a=2$ and $\lambda/\sigma=2$. The incoming speeds and impact parameters are (a) $g/\sqrt{\epsilon_a/m}=4$, $b/\lambda=0.3$; (b) $g/\sqrt{\epsilon_a/m}=4$, $b/\lambda=0.4$; (c) $g/\sqrt{\epsilon_a/m}=4$, $b/\lambda=0.8$; and (d) $g/\sqrt{\epsilon_a/m}={2}$, $b/\lambda=0.3$.}
\label{coll}
\end{figure}

It turns out that the projectile trajectories are equivalent to that of light rays traversing first a corona of width $(\lambda-\sigma)/2$ and (relative) refractive index  $n_a>1$, and then a spherical core of diameter $\sigma$ and (relative) refractive index  $n_r<1$, so that the laws of geometrical optics can be applied to obtain the scattering angle $\chi(b,g)$. If the  angle of incidence to the inner core is larger than the critical angle $\sin^{-1}(n_r/n_a)$ then total internal reflection exists and the ray does not penetrate into the core. The same may happen if $n_r$ is not a real quantity  (i.e., if $g<g_r^\thr$), in which case the core becomes opaque to the light ray, regardless of the angle of incidence.

\begin{table}[ht]
\caption{Classes of collisions depending on the relative speed and the impact parameter.}
\label{table:0}
\tabcolsep7pt\begin{tabular}{lllll}
\hline
PSW parameters&Relative speed&Impact parameter&Collision&$\chi(b,g)$\\
\hline
$\epsilon_r/\epsilon_a>\left(\lambda^2/\sigma^2-1\right)^{-1}$&$0<g<g_a^\thr$&$0<b<\lambda$&Hard&Eq.\ \eqref{2et14}\\
&$g_a^\thr<g<g_r^\thr$&$0<b<n_a(g)\sigma$&Hard&Eq.\ \eqref{2et14}\\
&&$n_a(g)\sigma<b<\lambda$&Grazing&Eq.\ \eqref{2et15}\\
&$g>g_r^\thr$&$0<b<n_r(g)\sigma$&Soft&Eq.\ \eqref{2et13}\\
&&$n_r(g)\sigma<b<n_a(g)\sigma$&Hard&Eq.\ \eqref{2et14}\\
&&$n_a(g)\sigma<b<\lambda$&Grazing&Eq.\ \eqref{2et15}\\
$\epsilon_r/\epsilon_a<\left(\lambda^2/\sigma^2-1\right)^{-1}$&$0<g<g_r^\thr$&$0<b<\lambda$&Hard&Eq.\ \eqref{2et14}\\
&$g_r^\thr<g<g_a^\thr$&$0<b<n_r(g)\sigma$&Soft&Eq.\ \eqref{2et13}\\
&&$n_r(g)\sigma<b<\lambda$&Hard&Eq.\ \eqref{2et14}\\
&$g>g_a^\thr$&$0<b<n_r(g)\sigma$&Soft&Eq.\ \eqref{2et13}\\
&&$n_r(g)\sigma<b<n_a(g)\sigma$&Hard&Eq.\ \eqref{2et14}\\
&&$n_a(g)\sigma<b<\lambda$&Grazing&Eq.\ \eqref{2et15}\\
\hline
\end{tabular}
\end{table}
\begin{figure}[ht]
\begin{tabular}{cc}
 \includegraphics[scale=0.4]{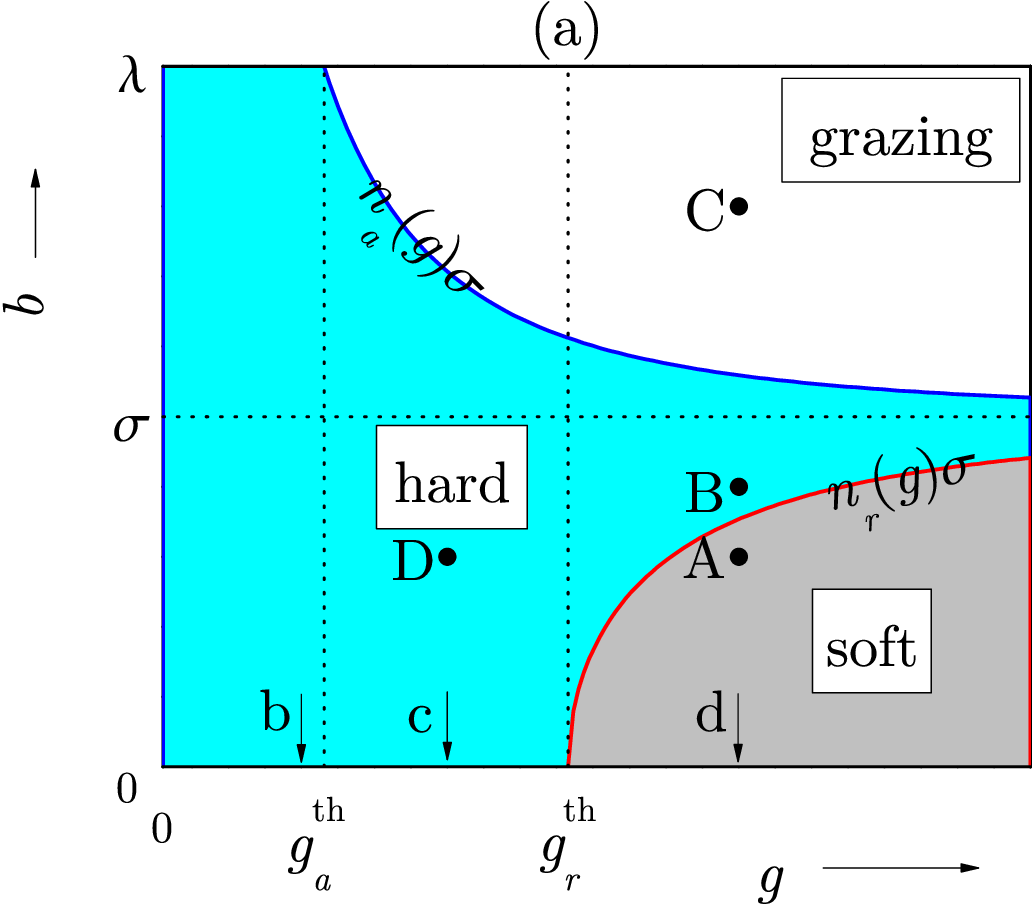}&\includegraphics[scale=0.7]{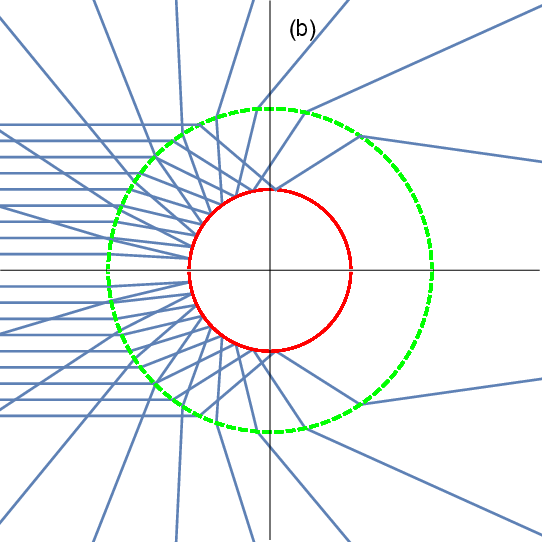}\\
\includegraphics[scale=0.7]{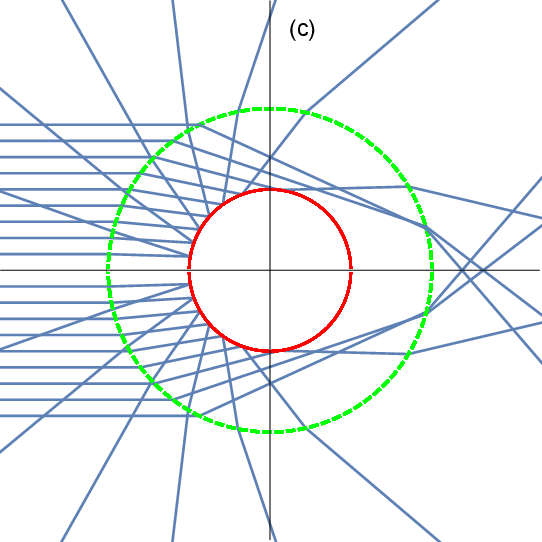}&\includegraphics[scale=0.7]{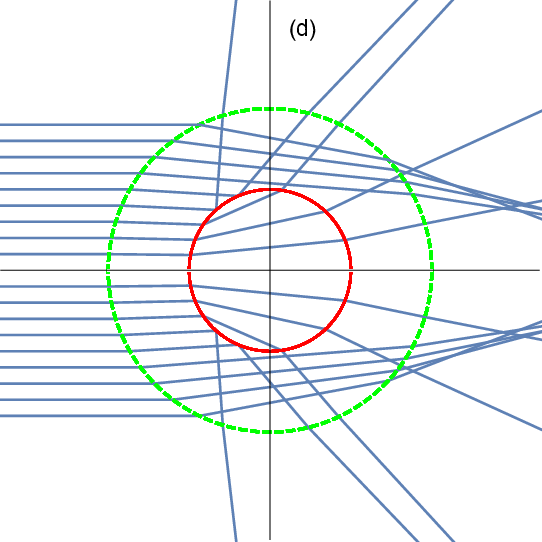}
\end{tabular}
\caption{(a) Representative ``phase diagram'' in the plane $b$ vs $g$ showing the regions where the collisions are soft, hard, or grazing. In this example, $\epsilon_r/\epsilon_a=2$ and $\lambda/\sigma=2$, as in Fig.~\ref{coll}; the points A--D correspond to the panels (a)--(d), respectively, of Fig.~\ref{coll}. (b) Trajectories with  $\epsilon_r/\epsilon_a=2$, $\lambda/\sigma=2$, and $g/\sqrt{\epsilon_a/m}=1$ [represented by the arrow labeled b in panel (a)]. (c) Trajectories with  $\epsilon_r/\epsilon_a=2$, $\lambda/\sigma=2$, and $g/\sqrt{\epsilon_a/m}={2}$ [represented by the arrow labeled c in panel (a)]. (d) Trajectories with  $\epsilon_r/\epsilon_a=2$, $\lambda/\sigma=2$, and $g/\sqrt{\epsilon_a/m}=4$ [represented by the arrow labeled d in panel (a)].}
\label{rays}
\end{figure}

\begin{figure}[ht]
\begin{tabular}{ccc}
 \includegraphics[scale=0.55]{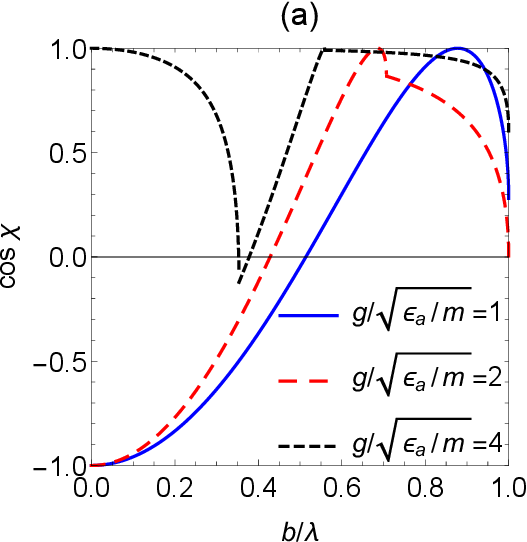}&\includegraphics[scale=0.55]{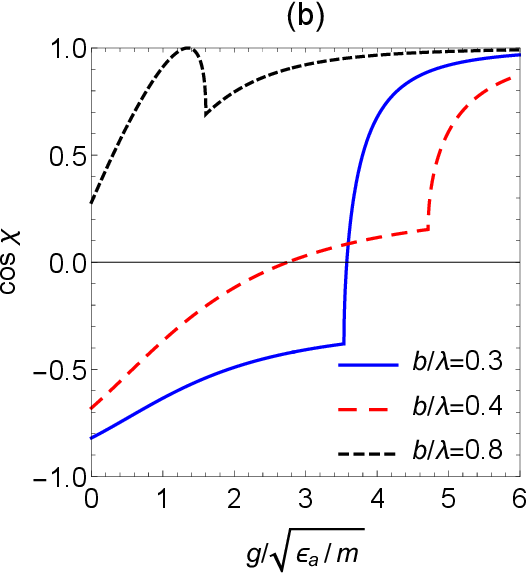}&\includegraphics[scale=0.52]{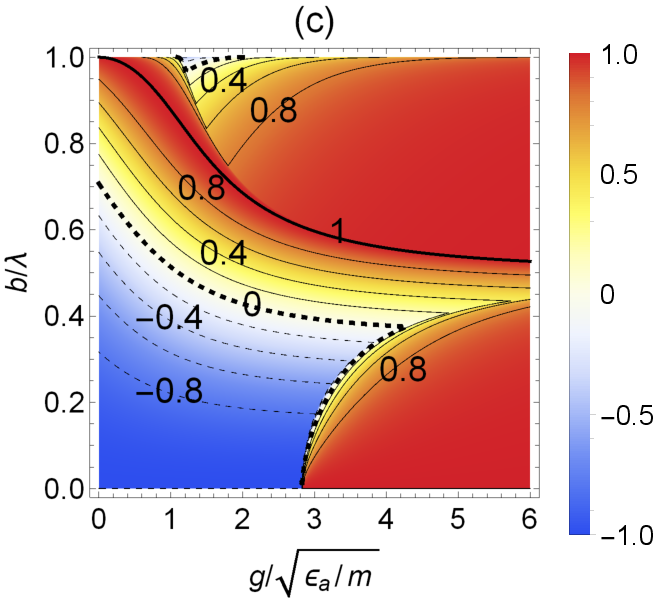}
\end{tabular}
\caption{Scattering angle for a PSW potential with $\epsilon_r/\epsilon_a=2$ and $\lambda/\sigma=2$, as in Figs.~\ref{coll} and \ref{rays}. (a) $\cos\chi(b,g)$ vs $b/\lambda$ for three representative values of $g/\sqrt{\epsilon_a/m}$. (b) $\cos\chi(b,g)$ vs $g/\sqrt{\epsilon_a/m}$ for three representative values of $b/\lambda$. (c) Density plot of $\cos\chi(b,g)$ in the plane $b/\lambda$ vs $g/\sqrt{\epsilon_a/m}$. The thin solid lines represent  points with $\cos\chi=0.8, 0.6, 0.4, 0.2$, while the thin dashed lines represent  points with $\cos\chi=-0.8, -0.6,-0.4, -0.2$. The two loci $\cos\chi=0$ are represented by thick dashed lines and the locus $\cos\chi=1$ is represented by a thick solid line.}
\label{coschi}
\end{figure}

There are then three possible classes of collisions: \emph{soft} collisions [see Fig.~\ref{coll}(a)],  \emph{hard} collisions [see Figs.~\ref{coll}(b) and \ref{coll}(d)], and  \emph{grazing} collisions [see Fig.~\ref{coll}(c)]. Let us first derive the scattering angle for soft collisions, which require $g>g_r^\thr$ and $b/\sigma<n_r(g)$. The angles of incidence ($\theta_i$, $\theta_i'$) and the angles of refraction  ($\theta_r$, $\theta_r'$) in Fig.~\ref{coll}(a) obey the relations
\begin{equation}
\sin{\theta_{i}}=\frac{b}{\lambda}, \quad \lambda\sin{\theta_{r}}=\sigma\sin{\theta_{i}'},\quad
\sin{\theta_{i}}=n_{a}(g)\sin{\theta_{r}}, \quad
n_{a}(g)\sin{\theta_{i}'}=n_{r}(g)\sin{\theta_{r}'},
\label{2et5}
\end{equation}
so that
\begin{equation}
\theta_{i}(b)=\sin^{-1}\frac{b}{\lambda}, \quad \theta_{r}(b,g)=\sin^{-1}\frac{b/\lambda}{n_a(g)},
\quad \theta_{i}'(b,g)=\sin^{-1}\frac{b/\sigma}{n_a(g)}, \quad
\theta_{r}'(b,g)=\sin^{-1}\frac{b/\sigma}{n_r(g)}.
\label{2et9}
\end{equation}
Therefore, in the case of soft collisions the scattering angle is [see Fig.~\ref{coll}(a)]
\begin{equation}
\chi(b,g)=\chi_{\text{soft}}(b,g)\equiv 2\left[ \theta_{r}(b,g)+\theta_{r}'(b,g)-\theta_{i}(b)-\theta_{i}'(b,g) \right]. \label{2et13}
\end{equation}

Even with $g>g_r^\thr$, if the impact parameter is larger than $ n_r(g)\sigma$, but smaller than both $ n_a(g)\sigma$ and $\lambda$, total internal reflection occurs and a hard collision takes place, as represented in Fig.~\ref{coll}(b). In such a case, the scattering angle is given by Eq.\ \eqref{2et13}, except for the formal change $\theta_{r}'\to\pi/2$, i.e.,
\begin{equation}
\chi(b,g)=\chi_{\text{hard}}(b,g)\equiv 2\left[ \theta_{r}(b,g)+\frac{\pi}{2}-\theta_{i}(b)-\theta_{i}'(b,g) \right]. \label{2et14}
\end{equation}

If $g>g_r^\thr$ and $ n_a(g)\sigma>\lambda$, only soft collisions (for $b<n_r\sigma$) and hard collisions (for $n_r\sigma<b<\lambda$) are possible. On the other hand, if $g$ is such that $ n_a(g)\sigma<\lambda$, i.e., if $g>g_a^\thr$, then a grazing collision occurs for impact parameters in the interval $n_a(g)\sigma<b<\lambda$, as represented in Fig.~\ref{coll}(c). In that case, the scattering angle is obtained from Eq.\ \eqref{2et14} by the formal change $\theta_{i}'\to\pi/2$, namely
\begin{equation}
\chi(b,g)=\chi_{\text{grazing}}(b,g)\equiv 2\left[ \theta_{r}(b,g)-\theta_{i}(b) \right]. \label{2et15}
\end{equation}

Thus far, we have assumed $g>g_r^{\text{th}}$. However, if  $g<g_r^{\text{th}}$, then soft collisions are absent. If, additionally,  $g<g_a^{\text{th}}$, all collisions are hard for any impact parameter. If $g_a^\thr<g<g_r^\thr$,  then collisions are hard for $b<n_{a}(g)\sigma$ [see Fig.~\ref{coll}(d) for an example] and grazing for $n_{a}(g)\sigma<b<\lambda$.

A summary of all the possible scenarios is presented in Table \ref{table:0}.
A representative ``phase diagram'' for a PSW potential with  $\epsilon_r/\epsilon_a>\left(\lambda^2/\sigma^2-1\right)^{-1}$ (so that $g_a^\thr<g_r^\thr$) is presented in Fig.~\ref{rays}(a). If $g<g_a^\thr$ all collisions are hard [see Fig.~\ref{rays}(b) for an example]. On the other hand, if $g_a^\thr<g<g_r^\thr$ collisions are hard for $b<n_a(g)\sigma$ and grazing for $n_a(g)\sigma<b<\lambda$ [see Fig.~\ref{rays}(c) for an example]. Finally, if $g>g_r^\thr$, collisions are soft for $b<n_r(g)\sigma$, hard for $n_r(g)\sigma<b<n_a(g)\sigma$, and grazing for $n_a(g)\sigma<b<\lambda$ [see Fig.~\ref{rays}(d) for an example]. In the case of a PSW potential with a narrow corona such that $\epsilon_r/\epsilon_a<\left(\lambda^2/\sigma^2-1\right)^{-1}$, the phase diagram would be qualitatively similar to that of Fig.~\ref{rays}(a), except that now $g_a^\thr>g_r^\thr$.
In the SW limit ($\epsilon_r\to\infty\Rightarrow g_r^\thr\to\infty$) the curve $n_r(g)\sigma$ in Fig.~\ref{rays}(a) would be absent and all the collisions would be either hard or grazing. In the opposite PS limit ($\epsilon_a\to0 \Rightarrow n_a\to 1$, all the collisions are either soft or hard; virtual grazing collisions with $\sigma<b<\lambda$ are actually null collisions. Therefore, we see that the PSW model (with finite $\epsilon_r$ and nonzero $\epsilon_a$) smoothly interpolates between the SW and PS models.

The dependence of $\cos\chi(b,g)$ on both the impact parameter $b$ and the relative speed $g$ is shown in Fig.~\ref{coschi} for the  PSW potential with $\epsilon_r/\epsilon_a=2$ and $\lambda/\sigma=2$, as in Figs.~\ref{coll} and \ref{rays}. Figure~\ref{coschi}(a) illustrates the dependence on $b$ for three representative values of $g$: a first value ($g/\sqrt{\epsilon_a/m}=1$) for which only hard collisions are possible, a second value ($g/\sqrt{\epsilon_a/m}=2$) for which collisions are hard or grazing, depending on  $b$, and a third value ($g/\sqrt{\epsilon_a/m}=4$) for which  collisions are soft, hard, or grazing. A kink in each curve signals the transition from soft to hard collisions or from hard to grazing collisions. For a more complete study, the reader can consult the interactive animation of Ref.\ \cite{note_18_09}. The dependence of $\cos\chi(b,g)$ on the reduced speed $g/\sqrt{\epsilon_a/m}$ is displayed in Fig.~\ref{coschi}(b) for three values of the impact parameter: for $b<\sigma$ ($b/\lambda=0.3$ and $b/\lambda=0.4$) the collisions change from hard to soft as $g$ increases, while for $b>\sigma$ ($b/\lambda=0.8$) the collisions change from hard to grazing with increasing  $g$ [see also Fig.~\ref{rays}(a)]. A global view of the dependence of $\cos\chi(b,g)$ on both $b$ and $g$ is offered by a density plot in Fig.~\ref{coschi}(c). The two loci $\cos\chi=0\Rightarrow \chi=\frac{\pi}{2}$  split the plane $b$ vs $g$ into a region of forward scattering ($\cos\chi>0\Rightarrow 0<\chi<\frac{\pi}{2}$) and two regions (one of them very small, close to $b/\lambda=1$) of backscattering ($\cos\chi<0\Rightarrow \frac{\pi}{2}<\chi<{\pi}$). Moreover, in the forward scattering region, the locus $\cos\chi=1\Rightarrow \chi=0$ defines (hard) scattering processes equivalent to null collisions.
Setting $\chi_{\text{hard}}=0$ in Eq.\ \eqref{2et14} and applying simple algebra one finds $\sin^2\theta_i+\sin^2\theta_i'+\sin^2\theta_r=1+2\sin\theta_i \sin\theta_i' \sin\theta_r$. Substitution of Eqs.\ \eqref{2et9} yields the following equation characterizing the locus $\cos\chi(b,g)=1$:
\beq
\left[1+n_a^2(g)+\frac{\lambda^2}{\sigma^2}\right]\frac{b^2}{\sigma^2}=
n_a^2(g)\frac{\lambda^2}{\sigma^2}+2\frac{{b}^3}{\sigma^3}.
\label{13}
\eeq

In the limit of high speeds ($g\to\infty$), the two ``refractive indices'' $n_a(g)$ and $n_r(g)$ tend to unity, so that the collisions (either soft, hard, or grazing) become null, i.e., $\lim_{g\to\infty}\cos\chi(b,g)=1$. In the opposite limit of small speeds ($g\to 0$), all the collisions are hard [see Fig.~\ref{rays}(a)] with $n_a(g)\approx 4\epsilon_a/mg^2\to\infty$. A series expansion of $\theta_r$ and $\theta_i'$ in powers of $n_a^{-1}$ yields
\beq
\cos\chi(b,g)\simeq 1-2\left[1-\frac{b^2}{\lambda^2}-\frac{b^2}{2\lambda}\left(\sigma^{-1}-\lambda^{-1}\right)\sqrt{1-\frac{b^2}{\lambda^2}}\frac{mg^2}{\epsilon_a}
+\mathcal{O}\left(g^{4}\right)\right]\Theta(\lambda-b),
\eeq
where $\Theta(x)$ is the Heaviside step function. Thus, in the limit $g\to 0$ the scattering angle is the same as that corresponding to hard spheres of diameter $\lambda$, i.e.,
$\lim_{g\to 0}\cos\chi(b,g)=1-2\left(1-b^2/\lambda^2\right)\Theta(\lambda-b)$.

\section{TRANSPORT COEFFICIENTS}
Once the scattering angle $\chi(b,g)$ has been determined, we are in conditions of obtaining the transport coefficients (shear viscosity, thermal conductivity, and self-diffusion coefficient) of a gas made of particles interacting via the PSW potential.

In general, the Chapman--Enskog method allows one to derive the Navier--Stokes transport coefficients from the Boltzmann equation for  a dilute gas   in terms of the scattering law corresponding to the interaction potential of interest \cite{CC70,FK72}.  In the first Sonine approximation,
\beq
\eta(T)=\frac{5}{8}\frac{k_BT}{\Omega_{2,2}(T)},\quad \kappa(T)=\frac{15}{4}\frac{k_B}{m}\eta(T),\quad D(T)=\frac{3}{8}\frac{k_BT}{mn\Omega_{1,1}(T)},
\label{6}
\eeq
where $\eta(T)$, $\kappa(T)$, and $D(T)$ are the shear viscosity, thermal conductivity, and self-diffusion coefficient, respectively. The collisional integrals $\Omega_{k,\ell}(T)$ are defined as
\beq
\Omega_{k,\ell}(T)\equiv\sqrt{\frac{k_BT}{\pi m}}\int_0^\infty dy\,e^{-y^2}y^{2k+3}Q_{\ell}\left(2y\sqrt{k_BT/m}\right),\quad Q_\ell(g)\equiv 2\pi \int_0^\infty db\, b\left[1-\cos^\ell\chi(b,g)\right].
\label{7}
\eeq
In the special case of hard spheres (HS) of diameter $\sigma$, one has $\cos\chi(b,g)=1-2\left(1-b^2/\sigma^2\right)\Theta(\sigma-b)$,  so that
\beq
\Omega_{k,\ell}^{\text{HS}}(T)=\sqrt{\frac{k_BT}{\pi m}}\pi\sigma^2(k+1)!\frac{1-(-1)^\ell+2\ell}{4(1+\ell)},
\label{n3}
\eeq
\beq
\eta_{\text{HS}}(T)=\frac{5}{16}\frac{\sqrt{mk_BT/\pi}}{\sigma^2},\quad \kappa_{\text{HS}}(T)=\frac{15}{4}\frac{k_B}{m}\eta_{\text{HS}}(T),\quad D_{\text{HS}}(T)=\frac{3}{8}\frac{\sqrt{k_BT/m\pi}}{n\sigma^2}.
\label{n2}
\eeq

Now we particularize to the PSW potential and introduce the
\textit{reduced} integrals $\Omega_{k,\ell}^*(T^*)=\Omega_{k,\ell}(T)/\Omega_{k,\ell}^{\text{HS}}(T)$, where $T^*$ is a reduced temperature. Since there are two energy scales ($\epsilon_r$ and $\epsilon_a$) in the PSW model, either of them can be used to scale the temperature. Here we choose the well depth $\epsilon_a$ to define $T^*=k_BT/\epsilon_a$, while we will denote by $T^\dagger=k_BT/\epsilon_r$ the other scaled temperature. Both quantities are simply related by $T^*=T^\dagger\epsilon_r/\epsilon_a$. The reduced transport coefficients are
\beq
\eta^*(T^*)\equiv\frac{\eta(T)}{\eta_{\text{HS}}(T)}=\frac{1}{\Omega_{2,2}^*(T^*)}, \quad
\kappa^*(T^*)\equiv \frac{\kappa(T)}{\kappa_{\text{HS}}(T)}=\frac{1}{\Omega_{2,2}^*(T^*)}, \quad
D^*(T^*)\equiv\frac{D(T)}{D_{\text{HS}}(T)}=\frac{1}{\Omega_{1,1}^*(T^*)}.
\label{9}
\eeq

Since in the limit $g\to 0$ the scattering angle is the same as that corresponding to hard spheres of diameter $\lambda$, while in the opposite limit $g\to\infty$ all the scattering processes tend to null collisions, one has the following forms in the low- and high-temperature limits,
\beq
\label{limitPSW}
\lim_{T^*\to 0}\Omega_{k,\ell}^*(T^*)=(\lambda/\sigma)^2,\quad  \lim_{T^*\to \infty}\Omega_{k,\ell}^*(T^*)=0,
\eeq
with independence of the value of $\epsilon_r/\epsilon_a$, provided it is finite. On the other hand in the SW ($\epsilon_r\to\infty$) and PS ($\epsilon_a\to 0$) models the  low- and high-temperature limits are
\begin{subequations}
\label{limitSW&PS}
\beq
\lim_{T^*\to 0}\Omega_{k,\ell}^{*,\text{SW}}(T^*)=(\lambda/\sigma)^2,\quad  \lim_{T^*\to \infty}\Omega_{k,\ell}^{*,\text{SW}}(T^*)=1,
\eeq
\beq
\lim_{T^\dagger\to 0}\Omega_{k,\ell}^{*,\text{PS}}(T^\dagger)=1,\quad  \lim_{T^\dagger\to \infty}\Omega_{k,\ell}^{*,\text{PS}}(T^\dagger)=0.
\eeq
\end{subequations}
Comparison between Eqs.\ \eqref{limitPSW} and \eqref{limitSW&PS} shows that the PSW model behaves as the SW model for low temperatures and as the PS model for high temperatures. The interesting question is to elucidate how the PSW model interpolates between those two opposite limits in the domain of moderate temperatures.

\begin{figure}[ht]
 \centerline{\includegraphics[scale=0.75]{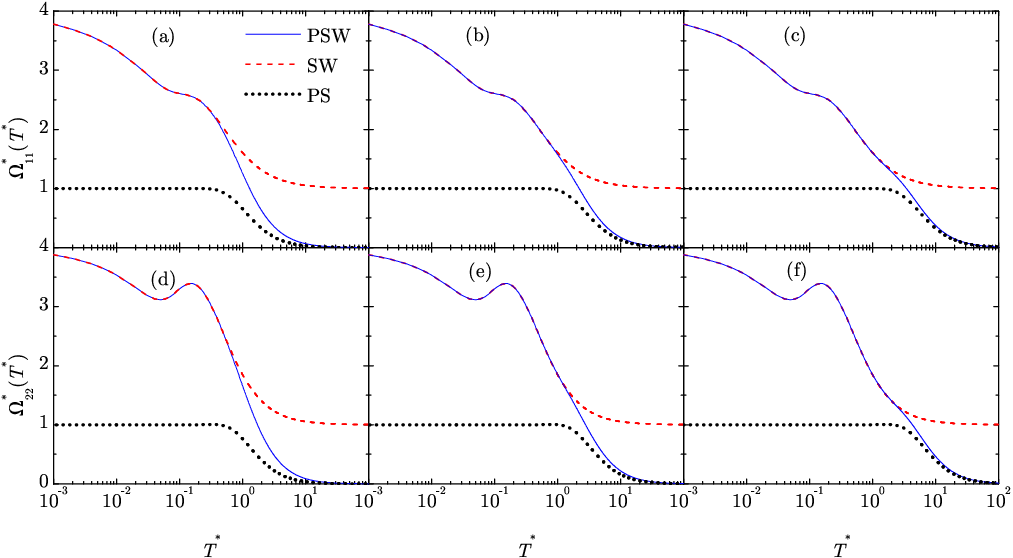}}
\caption{Plot of $\Omega_{11}^*(T^*)$ [panels (a)--(c)] and $\Omega_{22}^*(T^*)$ [panels (d)--(f)] vs $T^*=k_BT/\epsilon_a$ for the PSW model (solid lines), the SW model (dashed lines), and the PS model (dotted lines). In the PSW and SW models, $\lambda/\sigma=2$. Moreover, in the PSW model $\epsilon_r/\epsilon_a=2$ in panels (a) and (d), $\epsilon_r/\epsilon_a=5$ in panels (b) and (e), and $\epsilon_r/\epsilon_a=10$ in panels (c) and (f).
Since in the PS model the only reduced temperature is $T^\dagger=k_BT/\epsilon_r$, $T^*$ is defined for the PS curves as $T^*=2T^\dagger$ in panels (a) and (d), $T^*=5T^\dagger$ in panels (b) and (e), and $T^*=10T^\dagger$ in panels (c) and (f).}
\label{integrals}
\end{figure}

Figure~\ref{integrals} shows the (reduced) collisional integrals $\Omega_{11}^*(T^*)$  and $\Omega_{22}^*(T^*)$ for the PSW model with $\lambda/\sigma=2$ and $\epsilon_r/\epsilon_a=2$, $5$, and $10$. The curves corresponding to the SW model ($\epsilon_r\to\infty$) with the same size ratio $\lambda/\sigma=2$ are also included. In the case of the PS model ($\epsilon_a\to 0$) the quantity $T^*$ is meaningless and the relevant scaled temperature is $T^\dagger$. In order to compare the PSW and SW curves at common values of $T^\dagger$, we define a nominal $T^*$ for the PS model as $T^*=2T^\dagger$, $T^*=5T^\dagger$, and $T^*=10T^\dagger$, respectively.

From Fig.~\ref{integrals} we observe that the PSW curves (even in the case $\epsilon_r/\epsilon_a=2$)  are practically indistinguishable from the (common) SW curves for low enough temperatures ($T^\dagger<T^\dagger_{\text{SW}}\approx 0.2\Rightarrow T^*<T^*_{\text{SW}}\approx 0.2\epsilon_r/\epsilon_a$), and not just in the limit $T^*\to 0$. In particular, $\Omega_{11}^*(T^*)$  and $\Omega_{22}^*(T^*)$ start by rapidly decaying from the zero-temperature limit value $(\lambda/\sigma)^2=4$ as temperature increases; then, $\Omega_{11}^*(T^*)$ presents an inflection point at $T^*\approx 0.1$ (where $\Omega_{11}^*\approx 2.6$), while $\Omega_{22}^*(T^*)$ has a local minimum ($\Omega_{22}^*\approx 3.1$) at $T^*\approx 0.05$ followed by a local maximum  ($\Omega_{22}^*\approx 3.4$) at $T^*\approx 0.15$. For temperatures larger than about $T^*_{\text{SW}}\approx 0.2\epsilon_r/\epsilon_a$, the PSW curves separate from the SW ones,  decaying to zero for high temperatures in a way analogous to the PS curves. If one defines a typical temperature $T^*_{\text{PS}}=T^\dagger_{\text{PS}}\epsilon_r/\epsilon_a$  beyond which both $\Omega_{11}^*(T^*)$  and $\Omega_{22}^*(T^*)$ are smaller than about $0.02$ (i.e., $\eta^*$, $\kappa^*$, and $D^*$ are larger than about $50$), one finds that $T^\dagger_{\text{SW}}\approx 10$. Therefore, the \emph{transition} regime where the (reduced) PSW transport coefficients change from the SW values to very high values (comparable to those corresponding to the PS model) is $0.2\lesssim T^\dagger\lesssim 10$.

\section{APPLICATION TO THE PLANAR FOURIER FLOW}

From the shear viscosity and the thermal conductivity one can define an \textit{effective} collision frequency  as \cite{CC70}  $\nu(T)=nk_BT/\eta(T)=\frac{15}{4}nk_B^2T/m\kappa(T)$. Therefore,
\beq
\nu(T^*)=\nu_a\sqrt{T^*}\Omega_{2,2}^*(T^*),\quad \nu_a\equiv \frac{16}{5}n\sigma^2\sqrt{\pi\epsilon_a/m}.
\label{N19}
\eeq

Let us analyze now the temperature  profile in the steady planar Fourier flow, where the gas is enclosed between two parallel plates at rest located at $y=0$ and $y=L$, and kept at temperatures $T_1$ and $T_2$, respectively. It is known that in this geometry the Fourier law is very reliable, even for strong thermal gradients \cite{GS03,TCG82,SBG86,BSD87,MASG94,S09b}.  According to the Fourier law, the stationary temperature profile is the solution to $\nu^{-1}(T)\partial T/\partial y=\mbox{const}$. Neglecting boundary layer effects and applying the boundary conditions, the implicit solution is
\beq
y/L=\left[{\int_{T_1^*}^{T_2^*}d\theta\,\nu^{-1}(\theta)}\right]^{-1}{\int_{T_1^*}^{T^*}d\theta\,\nu^{-1}(\theta)}.
\label{N20}
\eeq
Given the values of $T_1^*$ and $T_2^*$, Eq.\ \eqref{N20} yields different temperature profiles depending on the values of $\lambda/\sigma$ and $\epsilon_r/\epsilon_a$.

\begin{figure}[ht]
\includegraphics[scale=0.75]{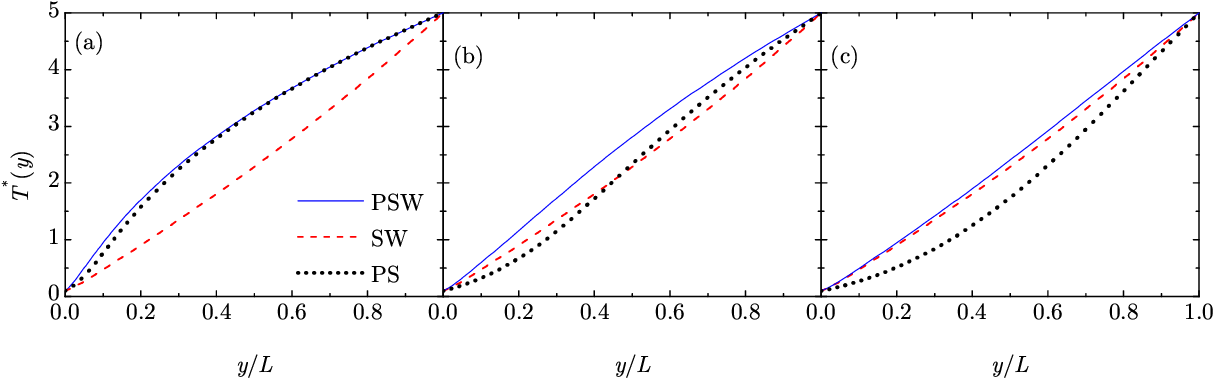}
\caption{Temperature profiles for the Fourier flow with $T_1^*=0.1$ and $T_2^*=5$ for the PSW model (solid lines), the SW model (dashed lines), and the PS model (dotted lines). In the PSW and SW models, $\lambda/\sigma=2$. Moreover, in the PSW model $\epsilon_r/\epsilon_a=2$ in panel (a), $\epsilon_r/\epsilon_a=5$ in panel (b), and $\epsilon_r/\epsilon_a=10$ in panel (c).
Since in the PS model the only reduced temperature is $T^\dagger=k_BT/\epsilon_r$, $T^*$ is defined for the PS curves as $T^*=2T^\dagger$ in panel (a), $T^*=5T^\dagger$ in panel (b), and $T^*=10T^\dagger$ in panel (c).
\label{profiles}}
\end{figure}

The temperature  profiles corresponding to the PS, SW, and PSW interaction models are compared in Fig.\ \ref{profiles} for the representative case $\lambda/\sigma=2$, $T_1^*=0.1$, $T_2^*=5$, and three values of the energy ratio: (a) $\epsilon_r/\epsilon_a=2$, (b) $\epsilon_r/\epsilon_a=5$, and (c) $\epsilon_r/\epsilon_a=10$. We observe that in cases (a) and (c) the PSW profiles are close to the PS and SW ones, respectively, while case (b) represents a transition situation where both attraction and penetrability are at stake.

\section{SUMMARY}
In this paper we have addressed the problem of determining the Navier--Stokes transport coefficients (shear viscosity, thermal conductivity, and self-diffusion coefficient) of a dilute gas made of particles interacting via the PSW potential. The potential is characterized by two dimensionless parameters: the diameter of the attractive corona relative to the diameter of the repulsive core ($\lambda/\sigma$) and the repulsive energy barrier relative to the attractive energy well ($\epsilon_r/\epsilon_a$). Since standard kinetic theory provides (approximate) formulas for the transport coefficients in terms of collision integrals involving the scattering angle $\chi(b,g)$ of a binary collision, we have first analyzed in detail the dependence of $\chi$ on both the impact parameter ($b$) and the relative speed ($g$). Three classes of collisions have been identified (soft, hard, and grazing), as described in Table \ref{table:0} and depicted in Figs.~\ref{coll} and \ref{rays}. In the case of the SW potential ($\epsilon_r\to\infty$), only hard and grazing collisions are present, while in the case of the PS potential ($\epsilon_a\to 0$), only soft and hard collisions exist.

By numerically performing the collision integrals \eqref{7}, their temperature dependence has been shown in Fig.~\ref{integrals} for $\lambda/\sigma=2$ and three representative values of $\epsilon_r/\epsilon_a$. As expected, the PSW transport coefficients smoothly interpolate between the SW and the PS ones. At a more specific level, it turns out that the PSW transport coefficients are practically indistinguishable from the SW ones if $k_BT\lesssim 0.2\epsilon_r$, while they are comparable to the PS ones if $k_BT\gtrsim 10\epsilon_r$. In the  transition regime,  $0.2\epsilon_r\lesssim k_BT\lesssim 10 \epsilon_r$, the values of the PSW transport coefficients result from an interplay between the two energy scales: while the transport coefficients are influenced by penetration effects (soft collisions), the attractive well also plays an important role (hard and grazing collisions).
As an illustration, we have analyzed the temperature profiles in the planar Fourier flow.

We expect that the results presented in this paper can stimulate the performance of computer simulations (by molecular dynamics or the direct simulation Monte Carlo method) to measure the transport properties of soft-matter potentials combining a penetrable core and an attractive tail. In this sense, our results can serve as a benchmark to assess the reliability of the simulation techniques in the dilute limit.

\section{ACKNOWLEDGMENTS}
The research of A.S. has been supported by the Spanish Agencia Estatal de Investigaci\'on through Grant No.\ FIS2016-76359-P
and the Junta de Extremadura (Spain) through Grant No.\ GR18079, both partially financed by Fondo Europeo de
Desarrollo Regional funds.

\nocite{*}

\begin{thebibliography}{1271}%
\makeatletter
\providecommand \@ifxundefined [1]{%
 \@ifx{#1\undefined}
}%
\providecommand \@ifnum [1]{%
 \ifnum #1\expandafter \@firstoftwo
 \else \expandafter \@secondoftwo
 \fi
}%
\providecommand \@ifx [1]{%
 \ifx #1\expandafter \@firstoftwo
 \else \expandafter \@secondoftwo
 \fi
}%
\providecommand \natexlab [1]{#1}%
\providecommand \enquote  [1]{``#1''}%
\providecommand \bibnamefont  [1]{#1}%
\providecommand \bibfnamefont [1]{#1}%
\providecommand \citenamefont [1]{#1}%
\providecommand \href@noop [0]{\@secondoftwo}%
\providecommand \href [0]{\begingroup \@sanitize@url \@href}%
\providecommand \@href[1]{\@@startlink{#1}\@@href}%
\providecommand \@@href[1]{\endgroup#1\@@endlink}%
\providecommand \@sanitize@url [0]{\catcode `\$12\catcode `\&12\catcode
  `\#12\catcode `\^12\catcode `\_12\catcode `\%12\relax}%
\providecommand \@@startlink[1]{}%
\providecommand \@@endlink[0]{}%
\providecommand \url  [0]{\begingroup\@sanitize@url \@url }%
\providecommand \@url [1]{\endgroup\@href {#1}{\urlprefix }}%
\providecommand \urlprefix  [0]{URL }%
\providecommand \Eprint [0]{\href }%
\providecommand \doibase [0]{http://dx.doi.org/}%
\providecommand \selectlanguage [0]{\@gobble}%
\providecommand \bibinfo  [0]{\@secondoftwo}%
\providecommand \bibfield  [0]{\@secondoftwo}%
\providecommand \translation [1]{[#1]}%
\providecommand \BibitemOpen [0]{}%
\providecommand \bibitemStop [0]{}%
\providecommand \bibitemNoStop [0]{.\EOS\space}%
\providecommand \EOS [0]{\spacefactor3000\relax}%
\providecommand \BibitemShut  [1]{\csname bibitem#1\endcsname}%
\let\auto@bib@innerbib\@empty
\bibitem [{\citenamefont {Chapman}\ and\ \citenamefont {Cowling}(1970)}]{CC70}%
  \BibitemOpen
  \bibfield  {author} {\bibinfo {author} {\bibfnamefont {S.}~\bibnamefont
  {Chapman}}\ and\ \bibinfo {author} {\bibfnamefont {T.~G.}\ \bibnamefont
  {Cowling}},\ }\href@noop {} {\emph {\bibinfo {title} {{The Mathematical
  Theory of Nonuniform Gases}}}}\ (\bibinfo  {publisher} {Cambridge University
  Press},\ \bibinfo {address} {Cambridge},\ \bibinfo {year} {1970})\BibitemShut
  {NoStop}%
\bibitem [{\citenamefont {Ferziger}\ and\ \citenamefont {Kaper}(1972)}]{FK72}%
  \BibitemOpen
  \bibfield  {author} {\bibinfo {author} {\bibfnamefont {J.~H.}\ \bibnamefont
  {Ferziger}}\ and\ \bibinfo {author} {\bibfnamefont {G.~H.}\ \bibnamefont
  {Kaper}},\ }\href@noop {} {\emph {\bibinfo {title} {{Mathematical Theory of
  Transport Processes in Gases}}}}\ (\bibinfo  {publisher} {North--Holland},\
  \bibinfo {address} {Amsterdam},\ \bibinfo {year} {1972})\BibitemShut
  {NoStop}%
\bibitem [{\citenamefont {Cercignani}(1988)}]{C88}%
  \BibitemOpen
  \bibfield  {author} {\bibinfo {author} {\bibfnamefont {C.}~\bibnamefont
  {Cercignani}},\ }\href@noop {} {\emph {\bibinfo {title} {{The Boltzmann
  Equation and Its Applications}}}}\ (\bibinfo  {publisher}
  {Springer--Verlag},\ \bibinfo {address} {New York},\ \bibinfo {year}
  {1988})\BibitemShut {NoStop}%
\bibitem [{\citenamefont {Garz{\'o}}\ and\ \citenamefont
  {Santos}(2003)}]{GS03}%
  \BibitemOpen
  \bibfield  {author} {\bibinfo {author} {\bibfnamefont {V.}~\bibnamefont
  {Garz{\'o}}}\ and\ \bibinfo {author} {\bibfnamefont {A.}~\bibnamefont
  {Santos}},\ }\href@noop {} {\emph {\bibinfo {title} {Kinetic Theory of Gases
  in Shear Flows: Nonlinear Transport}}},\ Fundamental Theories of Physics\
  (\bibinfo  {publisher} {Springer Netherlands},\ \bibinfo {year}
  {2003})\BibitemShut {NoStop}%
\bibitem [{\citenamefont {Cole}(1962)}]{C62}%
  \BibitemOpen
  \bibfield  {author} {\bibinfo {author} {\bibfnamefont {G.~H.~A.}\
  \bibnamefont {Cole}},\ }\href {\doibase 10.1063/1.1733352} {\bibfield
  {journal} {\bibinfo  {journal} {J. Chem. Phys.}\ }\textbf {\bibinfo {volume}
  {37}},\ \unskip\ \bibinfo {pages} {1631--1635} (\bibinfo {year}
  {1962})}\BibitemShut {NoStop}%
\bibitem [{\citenamefont {Marquest}\ and\ \citenamefont {Witten}(1989)}]{MW89}%
  \BibitemOpen
  \bibfield  {author} {\bibinfo {author} {\bibfnamefont {C.}~\bibnamefont
  {Marquest}}\ and\ \bibinfo {author} {\bibfnamefont {T.~A.}\ \bibnamefont
  {Witten}},\ }\href@noop {} {\bibfield  {journal} {\bibinfo  {journal} {J.
  Phys. France}\ }\textbf {\bibinfo {volume} {50}},\ \unskip\ \bibinfo {pages}
  {1267--1277} (\bibinfo {year} {1989})}\BibitemShut {NoStop}%
\bibitem [{\citenamefont {Stillinger}\ and\ \citenamefont
  {Stillinger}(1997)}]{SS97}%
  \BibitemOpen
  \bibfield  {author} {\bibinfo {author} {\bibfnamefont {F.~H.}\ \bibnamefont
  {Stillinger}}\ and\ \bibinfo {author} {\bibfnamefont {D.~K.}\ \bibnamefont
  {Stillinger}},\ }\href@noop {} {\bibfield  {journal} {\bibinfo  {journal}
  {Physica A}\ }\textbf {\bibinfo {volume} {244}},\ \unskip\ \bibinfo {pages}
  {358--369} (\bibinfo {year} {1997})}\BibitemShut {NoStop}%
\bibitem [{\citenamefont {Likos}, \citenamefont {Watzlawek},\ and\
  \citenamefont {L\"owen}(1998)}]{LWL98}%
  \BibitemOpen
  \bibfield  {author} {\bibinfo {author} {\bibfnamefont {C.~N.}\ \bibnamefont
  {Likos}}, \bibinfo {author} {\bibfnamefont {M.}~\bibnamefont {Watzlawek}}, \
  and\ \bibinfo {author} {\bibfnamefont {H.}~\bibnamefont {L\"owen}},\
  }\href@noop {} {\bibfield  {journal} {\bibinfo  {journal} {Phys. Rev. E}\
  }\textbf {\bibinfo {volume} {58}},\ \unskip\ \bibinfo {pages} {3135--3144}
  (\bibinfo {year} {1998})}\BibitemShut {NoStop}%
\bibitem [{\citenamefont {Likos}\ \emph {et~al.}(2001)\citenamefont {Likos},
  \citenamefont {Lang}, \citenamefont {Watzlawek},\ and\ \citenamefont
  {L\"owen}}]{LLWL01}%
  \BibitemOpen
  \bibfield  {author} {\bibinfo {author} {\bibfnamefont {C.~N.}\ \bibnamefont
  {Likos}}, \bibinfo {author} {\bibfnamefont {A.}~\bibnamefont {Lang}},
  \bibinfo {author} {\bibfnamefont {M.}~\bibnamefont {Watzlawek}}, \ and\
  \bibinfo {author} {\bibfnamefont {H.}~\bibnamefont {L\"owen}},\ }\href@noop
  {} {\bibfield  {journal} {\bibinfo  {journal} {Phys. Rev. E}\ }\textbf
  {\bibinfo {volume} {63}},\ p.\ \bibinfo {pages} {{031}{206}} (\bibinfo {year}
  {2001})}\BibitemShut {NoStop}%
\bibitem [{\citenamefont {Likos}(2001)}]{L01}%
  \BibitemOpen
  \bibfield  {author} {\bibinfo {author} {\bibfnamefont {C.~N.}\ \bibnamefont
  {Likos}},\ }\href {\doibase 10.1016/S0370-1573(00)00141-1} {\bibfield
  {journal} {\bibinfo  {journal} {Phys. Rep.}\ }\textbf {\bibinfo {volume}
  {348}},\ \unskip\ \bibinfo {pages} {267--439} (\bibinfo {year}
  {2001})}\BibitemShut {NoStop}%
\bibitem [{\citenamefont {Schmidt}\ and\ \citenamefont {Fuchs}(2002)}]{SF02}%
  \BibitemOpen
  \bibfield  {author} {\bibinfo {author} {\bibfnamefont {M.}~\bibnamefont
  {Schmidt}}\ and\ \bibinfo {author} {\bibfnamefont {M.}~\bibnamefont
  {Fuchs}},\ }\href@noop {} {\bibfield  {journal} {\bibinfo  {journal} {J.
  Chem. Phys.}\ }\textbf {\bibinfo {volume} {117}},\ \unskip\ \bibinfo {pages}
  {6308--6312} (\bibinfo {year} {2002})}\BibitemShut {NoStop}%
\bibitem [{\citenamefont {Acedo}\ and\ \citenamefont {Santos}(2004)}]{AS04}%
  \BibitemOpen
  \bibfield  {author} {\bibinfo {author} {\bibfnamefont {L.}~\bibnamefont
  {Acedo}}\ and\ \bibinfo {author} {\bibfnamefont {A.}~\bibnamefont {Santos}},\
  }\href@noop {} {\bibfield  {journal} {\bibinfo  {journal} {Phys. Lett. A}\
  }\textbf {\bibinfo {volume} {323}},\ \unskip\ \bibinfo {pages} {427--433}
  (\bibinfo {year} {2004})},\ \bibinfo {note} {erratum: \textbf{376},
  2274--2275 (2012)}\BibitemShut {NoStop}%
\bibitem [{\citenamefont {Malijevsk\'y}, \citenamefont {Yuste},\ and\
  \citenamefont {Santos}(2007)}]{MYS07}%
  \BibitemOpen
  \bibfield  {author} {\bibinfo {author} {\bibfnamefont {A.}~\bibnamefont
  {Malijevsk\'y}}, \bibinfo {author} {\bibfnamefont {S.~B.}\ \bibnamefont
  {Yuste}}, \ and\ \bibinfo {author} {\bibfnamefont {A.}~\bibnamefont
  {Santos}},\ }\href@noop {} {\bibfield  {journal} {\bibinfo  {journal} {Phys.
  Rev. E}\ }\textbf {\bibinfo {volume} {76}},\ p.\ \bibinfo {pages}
  {{021}{504}} (\bibinfo {year} {2007})}\BibitemShut {NoStop}%
\bibitem [{\citenamefont {Prestipino}\ \emph {et~al.}(2014)\citenamefont
  {Prestipino}, \citenamefont {Speranza}, \citenamefont {Malescio},\ and\
  \citenamefont {Giaquinta}}]{PSMG14}%
  \BibitemOpen
  \bibfield  {author} {\bibinfo {author} {\bibfnamefont {S.}~\bibnamefont
  {Prestipino}}, \bibinfo {author} {\bibfnamefont {C.}~\bibnamefont
  {Speranza}}, \bibinfo {author} {\bibfnamefont {G.}~\bibnamefont {Malescio}},
  \ and\ \bibinfo {author} {\bibfnamefont {P.~V.}\ \bibnamefont {Giaquinta}},\
  }\href {\doibase 10.1063/1.4866897} {\bibfield  {journal} {\bibinfo
  {journal} {J. Chem. Phys.}\ }\textbf {\bibinfo {volume} {140}},\ p.\ \bibinfo
  {pages} {084906} (\bibinfo {year} {2014})}\BibitemShut {NoStop}%
\bibitem [{\citenamefont {Malescio}\ and\ \citenamefont
  {Prestipino}(2015)}]{MP15}%
  \BibitemOpen
  \bibfield  {author} {\bibinfo {author} {\bibfnamefont {G.}~\bibnamefont
  {Malescio}}\ and\ \bibinfo {author} {\bibfnamefont {S.}~\bibnamefont
  {Prestipino}},\ }\href {\doibase 10.1103/PhysRevE.92.050301} {\bibfield
  {journal} {\bibinfo  {journal} {Phys. Rev. E}\ }\textbf {\bibinfo {volume}
  {92}},\ p.\ \bibinfo {pages} {050301} (\bibinfo {year} {2015})}\BibitemShut
  {NoStop}%
\bibitem [{\citenamefont {Malescio}, \citenamefont {Parola},\ and\
  \citenamefont {Prestipino}(2018)}]{MPP18}%
  \BibitemOpen
  \bibfield  {author} {\bibinfo {author} {\bibfnamefont {G.}~\bibnamefont
  {Malescio}}, \bibinfo {author} {\bibfnamefont {A.}~\bibnamefont {Parola}}, \
  and\ \bibinfo {author} {\bibfnamefont {S.}~\bibnamefont {Prestipino}},\
  }\href {\doibase 10.1063/1.5017566} {\bibfield  {journal} {\bibinfo
  {journal} {J. Chem. Phys.}\ }\textbf {\bibinfo {volume} {148}},\ p.\ \bibinfo
  {pages} {084904} (\bibinfo {year} {2018})}\BibitemShut {NoStop}%
\bibitem [{\citenamefont {Altenberger}(1975)}]{A75}%
  \BibitemOpen
  \bibfield  {author} {\bibinfo {author} {\bibfnamefont {A.~R.}\ \bibnamefont
  {Altenberger}},\ }\href {\doibase
  https://doi.org/10.1016/0378-4371(75)90145-4} {\bibfield  {journal} {\bibinfo
   {journal} {Physica A}\ }\textbf {\bibinfo {volume} {80}},\ \unskip\ \bibinfo
  {pages} {46--62} (\bibinfo {year} {1975})}\BibitemShut {NoStop}%
\bibitem [{\citenamefont {Groome}, \citenamefont {Dufty},\ and\ \citenamefont
  {Lindenfeld}(1979)}]{GDL79}%
  \BibitemOpen
  \bibfield  {author} {\bibinfo {author} {\bibfnamefont {L.}~\bibnamefont
  {Groome}}, \bibinfo {author} {\bibfnamefont {J.~W.}\ \bibnamefont {Dufty}}, \
  and\ \bibinfo {author} {\bibfnamefont {M.~J.}\ \bibnamefont {Lindenfeld}},\
  }\href {\doibase 10.1103/PhysRevA.19.304} {\bibfield  {journal} {\bibinfo
  {journal} {Phys. Rev. A}\ }\textbf {\bibinfo {volume} {19}},\ \unskip\
  \bibinfo {pages} {304--323} (\bibinfo {year} {1979})}\BibitemShut {NoStop}%
\bibitem [{\citenamefont {Santos}(2005)}]{S05b}%
  \BibitemOpen
  \bibfield  {author} {\bibinfo {author} {\bibfnamefont {A.}~\bibnamefont
  {Santos}},\ }\href {\doibase 10.1063/1.1941550} {\bibfield  {journal}
  {\bibinfo  {journal} {AIP Conf. Proc.}\ }\textbf {\bibinfo {volume} {762}},\
  \unskip\ \bibinfo {pages} {276--281} (\bibinfo {year} {2005})}\BibitemShut
  {NoStop}%
\bibitem [{\citenamefont {Mausbach}\ and\ \citenamefont {May}(2006)}]{MM06}%
  \BibitemOpen
  \bibfield  {author} {\bibinfo {author} {\bibfnamefont {P.}~\bibnamefont
  {Mausbach}}\ and\ \bibinfo {author} {\bibfnamefont {H.-O.}\ \bibnamefont
  {May}},\ }\href {\doibase 0.1524/zpch.2009.6056} {\bibfield  {journal}
  {\bibinfo  {journal} {Fluid Phase Equil.}\ }\textbf {\bibinfo {volume}
  {223}},\ \unskip\ \bibinfo {pages} {1035--1046} (\bibinfo {year}
  {2006})}\BibitemShut {NoStop}%
\bibitem [{\citenamefont {May}\ and\ \citenamefont {Mausbach}(2007)}]{MM07}%
  \BibitemOpen
  \bibfield  {author} {\bibinfo {author} {\bibfnamefont {H.-O.}\ \bibnamefont
  {May}}\ and\ \bibinfo {author} {\bibfnamefont {P.}~\bibnamefont {Mausbach}},\
  }\href {\doibase 10.1103/PhysRevE.76.031201} {\bibfield  {journal} {\bibinfo
  {journal} {Phys. Rev. E}\ }\textbf {\bibinfo {volume} {76}},\ p.\ \bibinfo
  {pages} {031201} (\bibinfo {year} {2007})}\BibitemShut {NoStop}%
\bibitem [{\citenamefont {Mausbach}\ and\ \citenamefont {May}(2009)}]{MM09}%
  \BibitemOpen
  \bibfield  {author} {\bibinfo {author} {\bibfnamefont {P.}~\bibnamefont
  {Mausbach}}\ and\ \bibinfo {author} {\bibfnamefont {H.-O.}\ \bibnamefont
  {May}},\ }\href {\doibase 10.1016/j.fluid.2006.07.021} {\bibfield  {journal}
  {\bibinfo  {journal} {Z. Phys. Chem.}\ }\textbf {\bibinfo {volume} {249}},\
  \unskip\ \bibinfo {pages} {17--23} (\bibinfo {year} {2009})}\BibitemShut
  {NoStop}%
\bibitem [{\citenamefont {Ahmed}, \citenamefont {Mausbach},\ and\ \citenamefont
  {Sadus}(2009)}]{AMS09}%
  \BibitemOpen
  \bibfield  {author} {\bibinfo {author} {\bibfnamefont {A.}~\bibnamefont
  {Ahmed}}, \bibinfo {author} {\bibfnamefont {P.}~\bibnamefont {Mausbach}}, \
  and\ \bibinfo {author} {\bibfnamefont {R.~J.}\ \bibnamefont {Sadus}},\ }\href
  {\doibase 10.1063/1.3273083} {\bibfield  {journal} {\bibinfo  {journal} {J.
  Chem. Phys.}\ }\textbf {\bibinfo {volume} {131}},\ p.\ \bibinfo {pages}
  {224511} (\bibinfo {year} {2009})}\BibitemShut {NoStop}%
\bibitem [{\citenamefont {Krekelberg}\ \emph {et~al.}(2009)\citenamefont
  {Krekelberg}, \citenamefont {Kumar}, \citenamefont {Mittal}, \citenamefont
  {Errington},\ and\ \citenamefont {Truskett}}]{KKMET09}%
  \BibitemOpen
  \bibfield  {author} {\bibinfo {author} {\bibfnamefont {W.~P.}\ \bibnamefont
  {Krekelberg}}, \bibinfo {author} {\bibfnamefont {T.}~\bibnamefont {Kumar}},
  \bibinfo {author} {\bibfnamefont {J.}~\bibnamefont {Mittal}}, \bibinfo
  {author} {\bibfnamefont {J.~R.}\ \bibnamefont {Errington}}, \ and\ \bibinfo
  {author} {\bibfnamefont {T.~M.}\ \bibnamefont {Truskett}},\ }\href {\doibase
  10.1103/PhysRevE.79.031203} {\bibfield  {journal} {\bibinfo  {journal} {Phys.
  Rev. E}\ }\textbf {\bibinfo {volume} {79}},\ p.\ \bibinfo {pages} {031203}
  (\bibinfo {year} {2009})}\BibitemShut {NoStop}%
\bibitem [{\citenamefont {Pond}\ \emph {et~al.}(2009)\citenamefont {Pond},
  \citenamefont {Krekelberg}, \citenamefont {Shen}, \citenamefont {Errington},\
  and\ \citenamefont {Truskett}}]{PKSET09}%
  \BibitemOpen
  \bibfield  {author} {\bibinfo {author} {\bibfnamefont {M.~J.}\ \bibnamefont
  {Pond}}, \bibinfo {author} {\bibfnamefont {W.~P.}\ \bibnamefont
  {Krekelberg}}, \bibinfo {author} {\bibfnamefont {V.~K.}\ \bibnamefont
  {Shen}}, \bibinfo {author} {\bibfnamefont {J.~R.}\ \bibnamefont {Errington}},
  \ and\ \bibinfo {author} {\bibfnamefont {T.~M.}\ \bibnamefont {Truskett}},\
  }\href {\doibase 10.1063/1.3256235} {\bibfield  {journal} {\bibinfo
  {journal} {J. Chem. Phys.}\ }\textbf {\bibinfo {volume} {131}},\ p.\ \bibinfo
  {pages} {161101} (\bibinfo {year} {2009})}\BibitemShut {NoStop}%
\bibitem [{\citenamefont {Shall}\ and\ \citenamefont {Egorov}(2010)}]{SE10}%
  \BibitemOpen
  \bibfield  {author} {\bibinfo {author} {\bibfnamefont {L.~A.}\ \bibnamefont
  {Shall}}\ and\ \bibinfo {author} {\bibfnamefont {S.~A.}\ \bibnamefont
  {Egorov}},\ }\href {\doibase 10.1063/1.3429354} {\bibfield  {journal}
  {\bibinfo  {journal} {J. Chem. Phys.}\ }\textbf {\bibinfo {volume} {132}},\
  p.\ \bibinfo {pages} {184504} (\bibinfo {year} {2010})}\BibitemShut {NoStop}%
\bibitem [{\citenamefont {Suh}\ \emph {et~al.}(2010)\citenamefont {Suh},
  \citenamefont {Kim}, \citenamefont {Kim},\ and\ \citenamefont
  {Santos}}]{SKKS10}%
  \BibitemOpen
  \bibfield  {author} {\bibinfo {author} {\bibfnamefont {S.-H.}\ \bibnamefont
  {Suh}}, \bibinfo {author} {\bibfnamefont {C.-H.}\ \bibnamefont {Kim}},
  \bibinfo {author} {\bibfnamefont {S.-C.}\ \bibnamefont {Kim}}, \ and\
  \bibinfo {author} {\bibfnamefont {A.}~\bibnamefont {Santos}},\ }\href
  {\doibase 10.1103/PhysRevE.82.051202} {\bibfield  {journal} {\bibinfo
  {journal} {Phys. Rev. E}\ }\textbf {\bibinfo {volume} {82}},\ p.\ \bibinfo
  {pages} {051202} (\bibinfo {year} {2010})}\BibitemShut {NoStop}%
\bibitem [{\citenamefont {Santos}, \citenamefont {Fantoni},\ and\ \citenamefont
  {Giacometti}(2008)}]{SFG08}%
  \BibitemOpen
  \bibfield  {author} {\bibinfo {author} {\bibfnamefont {A.}~\bibnamefont
  {Santos}}, \bibinfo {author} {\bibfnamefont {R.}~\bibnamefont {Fantoni}}, \
  and\ \bibinfo {author} {\bibfnamefont {A.}~\bibnamefont {Giacometti}},\
  }\href@noop {} {\bibfield  {journal} {\bibinfo  {journal} {Phys. Rev. E}\
  }\textbf {\bibinfo {volume} {77}},\ p.\ \bibinfo {pages} {{051}{206}}
  (\bibinfo {year} {2008})}\BibitemShut {NoStop}%
\bibitem [{\citenamefont {Fantoni}\ \emph {et~al.}(2009)\citenamefont
  {Fantoni}, \citenamefont {Giacometti}, \citenamefont {Malijevsk\'y},\ and\
  \citenamefont {Santos}}]{FGMS09}%
  \BibitemOpen
  \bibfield  {author} {\bibinfo {author} {\bibfnamefont {R.}~\bibnamefont
  {Fantoni}}, \bibinfo {author} {\bibfnamefont {A.}~\bibnamefont {Giacometti}},
  \bibinfo {author} {\bibfnamefont {A.}~\bibnamefont {Malijevsk\'y}}, \ and\
  \bibinfo {author} {\bibfnamefont {A.}~\bibnamefont {Santos}},\ }\href@noop {}
  {\bibfield  {journal} {\bibinfo  {journal} {J. Chem. Phys.}\ }\textbf
  {\bibinfo {volume} {131}},\ p.\ \bibinfo {pages} {{124}{106}} (\bibinfo
  {year} {2009})}\BibitemShut {NoStop}%
\bibitem [{\citenamefont {Fantoni}\ \emph {et~al.}(2010)\citenamefont
  {Fantoni}, \citenamefont {Giacometti}, \citenamefont {Malijevsk\'y},\ and\
  \citenamefont {Santos}}]{FGMS10}%
  \BibitemOpen
  \bibfield  {author} {\bibinfo {author} {\bibfnamefont {R.}~\bibnamefont
  {Fantoni}}, \bibinfo {author} {\bibfnamefont {A.}~\bibnamefont {Giacometti}},
  \bibinfo {author} {\bibfnamefont {A.}~\bibnamefont {Malijevsk\'y}}, \ and\
  \bibinfo {author} {\bibfnamefont {A.}~\bibnamefont {Santos}},\ }\href@noop {}
  {\bibfield  {journal} {\bibinfo  {journal} {J. Chem. Phys.}\ }\textbf
  {\bibinfo {volume} {133}},\ p.\ \bibinfo {pages} {{024}{101}} (\bibinfo
  {year} {2010})}\BibitemShut {NoStop}%
\bibitem [{\citenamefont {Fantoni}\ \emph
  {et~al.}(2011{\natexlab{a}})\citenamefont {Fantoni}, \citenamefont
  {Malijevsk\'y}, \citenamefont {Santos},\ and\ \citenamefont
  {Giacometti}}]{FMSG11a}%
  \BibitemOpen
  \bibfield  {author} {\bibinfo {author} {\bibfnamefont {R.}~\bibnamefont
  {Fantoni}}, \bibinfo {author} {\bibfnamefont {A.}~\bibnamefont
  {Malijevsk\'y}}, \bibinfo {author} {\bibfnamefont {A.}~\bibnamefont
  {Santos}}, \ and\ \bibinfo {author} {\bibfnamefont {A.}~\bibnamefont
  {Giacometti}},\ }\href {\doibase 10.1209/0295-5075/93/26002} {\bibfield
  {journal} {\bibinfo  {journal} {EPL}\ }\textbf {\bibinfo {volume} {93}},\ p.\
  \bibinfo {pages} {26002} (\bibinfo {year} {2011}{\natexlab{a}})}\BibitemShut
  {NoStop}%
\bibitem [{\citenamefont {Fantoni}\ \emph
  {et~al.}(2011{\natexlab{b}})\citenamefont {Fantoni}, \citenamefont
  {Malijevsk\'y}, \citenamefont {Santos},\ and\ \citenamefont
  {Giacometti}}]{FMSG11b}%
  \BibitemOpen
  \bibfield  {author} {\bibinfo {author} {\bibfnamefont {R.}~\bibnamefont
  {Fantoni}}, \bibinfo {author} {\bibfnamefont {A.}~\bibnamefont
  {Malijevsk\'y}}, \bibinfo {author} {\bibfnamefont {A.}~\bibnamefont
  {Santos}}, \ and\ \bibinfo {author} {\bibfnamefont {A.}~\bibnamefont
  {Giacometti}},\ }\href {\doibase 10.1080/00268976.2011.597357} {\bibfield
  {journal} {\bibinfo  {journal} {Mol. Phys.}\ }\textbf {\bibinfo {volume}
  {109}},\ \unskip\ \bibinfo {pages} {2723--2736} (\bibinfo {year}
  {2011}{\natexlab{b}})}\BibitemShut {NoStop}%
\bibitem [{\citenamefont {Malescio}(2014)}]{M14}%
  \BibitemOpen
  \bibfield  {author} {\bibinfo {author} {\bibfnamefont {G.}~\bibnamefont
  {Malescio}},\ }\href {\doibase 10.1080/00268976.2013.860246} {\bibfield
  {journal} {\bibinfo  {journal} {Mol. Phys.}\ }\textbf {\bibinfo {volume}
  {112}},\ \unskip\ \bibinfo {pages} {1731--1735} (\bibinfo {year}
  {2014})}\BibitemShut {NoStop}%
\bibitem [{\citenamefont {Holleran}\ and\ \citenamefont
  {Hulburt}(1951)}]{HH51}%
  \BibitemOpen
  \bibfield  {author} {\bibinfo {author} {\bibfnamefont {E.~M.}\ \bibnamefont
  {Holleran}}\ and\ \bibinfo {author} {\bibfnamefont {H.~M.}\ \bibnamefont
  {Hulburt}},\ }\href {\doibase 10.1063/1.1748167} {\bibfield  {journal}
  {\bibinfo  {journal} {J. Chem. Phys.}\ }\textbf {\bibinfo {volume} {19}},\
  \unskip\ \bibinfo {pages} {232--241} (\bibinfo {year} {1951})}\BibitemShut
  {NoStop}%
\bibitem [{\citenamefont {Longuet-Higgins}\ and\ \citenamefont
  {Valleau}(1958)}]{LHV58}%
  \BibitemOpen
  \bibfield  {author} {\bibinfo {author} {\bibfnamefont {H.~C.}\ \bibnamefont
  {Longuet-Higgins}}\ and\ \bibinfo {author} {\bibfnamefont {J.~P.}\
  \bibnamefont {Valleau}},\ }\href {\doibase 10.1080/00268975800100331}
  {\bibfield  {journal} {\bibinfo  {journal} {Mol. Phys.}\ }\textbf {\bibinfo
  {volume} {1}},\ \unskip\ \bibinfo {pages} {284--294} (\bibinfo {year}
  {1958})}\BibitemShut {NoStop}%
\bibitem [{\citenamefont {Hirschfelder}, \citenamefont {Curtiss},\ and\
  \citenamefont {Bird}(1964)}]{HCB64}%
  \BibitemOpen
  \bibfield  {author} {\bibinfo {author} {\bibfnamefont {J.~O.}\ \bibnamefont
  {Hirschfelder}}, \bibinfo {author} {\bibfnamefont {C.~F.}\ \bibnamefont
  {Curtiss}}, \ and\ \bibinfo {author} {\bibfnamefont {R.~B.}\ \bibnamefont
  {Bird}},\ }\href@noop {} {\emph {\bibinfo {title} {{Molecular Theory of Gases
  and Liquids}}}}\ (\bibinfo  {publisher} {Wiley},\ \bibinfo {address} {New
  York},\ \bibinfo {year} {1964})\BibitemShut {NoStop}%
\bibitem [{\citenamefont {Karkheck}\ and\ \citenamefont {Stell}(1983)}]{KS83}%
  \BibitemOpen
  \bibfield  {author} {\bibinfo {author} {\bibfnamefont {J.}~\bibnamefont
  {Karkheck}}\ and\ \bibinfo {author} {\bibfnamefont {G.}~\bibnamefont
  {Stell}},\ }\href@noop {} {\bibfield  {journal} {\bibinfo  {journal} {J.
  Phys. Chem.}\ }\textbf {\bibinfo {volume} {87}},\ \unskip\ \bibinfo {pages}
  {2858--2866} (\bibinfo {year} {1983})}\BibitemShut {NoStop}%
\bibitem [{\citenamefont {Korlipara}, \citenamefont {Stell},\ and\
  \citenamefont {Karkheck}(1989)}]{KSK89}%
  \BibitemOpen
  \bibfield  {author} {\bibinfo {author} {\bibfnamefont {R.~K.}\ \bibnamefont
  {Korlipara}}, \bibinfo {author} {\bibfnamefont {G.}~\bibnamefont {Stell}}, \
  and\ \bibinfo {author} {\bibfnamefont {J.}~\bibnamefont {Karkheck}},\
  }\href@noop {} {\bibfield  {journal} {\bibinfo  {journal} {J. Chem. Phys.}\
  }\textbf {\bibinfo {volume} {90}},\ \unskip\ \bibinfo {pages} {5687--5695}
  (\bibinfo {year} {1989})}\BibitemShut {NoStop}%
\bibitem [{\citenamefont {Pajuelo}\ and\ \citenamefont
  {Santos}(2011{\natexlab{a}})}]{note_18_09}%
  \BibitemOpen
  \bibfield  {author} {\bibinfo {author} {\bibfnamefont {P.}~\bibnamefont
  {Pajuelo}}\ and\ \bibinfo {author} {\bibfnamefont {A.}~\bibnamefont
  {Santos}},\ }\href@noop {} {}\  \bibinfo {year} {2011}{\natexlab{a}}
  \unskip,\ \bibinfo {note} {``Classical Scattering with a Penetrable
  Square-Well Potential'', Wolfram Demonstrations Project,\\
  http://demonstrations.wolfram.com/ClassicalScatteringWithAPenetrableSquareWellPotential/}\BibitemShut
  {NoStop}%
\bibitem [{\citenamefont {Tenenbaum}, \citenamefont {Ciccotti},\ and\
  \citenamefont {Gallico}(1982)}]{TCG82}%
  \BibitemOpen
  \bibfield  {author} {\bibinfo {author} {\bibfnamefont {A.}~\bibnamefont
  {Tenenbaum}}, \bibinfo {author} {\bibfnamefont {G.}~\bibnamefont {Ciccotti}},
  \ and\ \bibinfo {author} {\bibfnamefont {R.}~\bibnamefont {Gallico}},\ }\href
  {\doibase 10.1103/PhysRevA.25.2778} {\bibfield  {journal} {\bibinfo
  {journal} {Phys. Rev. A}\ }\textbf {\bibinfo {volume} {25}},\ \unskip\
  \bibinfo {pages} {2778--2787} (\bibinfo {year} {1982})}\BibitemShut {NoStop}%
\bibitem [{\citenamefont {Santos}, \citenamefont {Brey},\ and\ \citenamefont
  {Garz\'o}(1986)}]{SBG86}%
  \BibitemOpen
  \bibfield  {author} {\bibinfo {author} {\bibfnamefont {A.}~\bibnamefont
  {Santos}}, \bibinfo {author} {\bibfnamefont {J.~J.}\ \bibnamefont {Brey}}, \
  and\ \bibinfo {author} {\bibfnamefont {V.}~\bibnamefont {Garz\'o}},\ }\href
  {\doibase 10.1103/PhysRevA.34.5047} {\bibfield  {journal} {\bibinfo
  {journal} {Phys. Rev. A}\ }\textbf {\bibinfo {volume} {34}},\ \unskip\
  \bibinfo {pages} {5047--5050} (\bibinfo {year} {1986})}\BibitemShut {NoStop}%
\bibitem [{\citenamefont {Brey}, \citenamefont {Santos},\ and\ \citenamefont
  {Dufty}(1987)}]{BSD87}%
  \BibitemOpen
  \bibfield  {author} {\bibinfo {author} {\bibfnamefont {J.~J.}\ \bibnamefont
  {Brey}}, \bibinfo {author} {\bibfnamefont {A.}~\bibnamefont {Santos}}, \ and\
  \bibinfo {author} {\bibfnamefont {J.~W.}\ \bibnamefont {Dufty}},\ }\href
  {\doibase 10.1103/PhysRevA.34.5047} {\bibfield  {journal} {\bibinfo
  {journal} {Phys. Rev. A}\ }\textbf {\bibinfo {volume} {36}},\ \unskip\
  \bibinfo {pages} {2842--2849} (\bibinfo {year} {1987})}\BibitemShut {NoStop}%
\bibitem [{\citenamefont {Montanero}\ \emph {et~al.}(1994)\citenamefont
  {Montanero}, \citenamefont {Alaoui}, \citenamefont {Santos},\ and\
  \citenamefont {Garz\'o}}]{MASG94}%
  \BibitemOpen
  \bibfield  {author} {\bibinfo {author} {\bibfnamefont {J.~M.}\ \bibnamefont
  {Montanero}}, \bibinfo {author} {\bibfnamefont {M.}~\bibnamefont {Alaoui}},
  \bibinfo {author} {\bibfnamefont {A.}~\bibnamefont {Santos}}, \ and\ \bibinfo
  {author} {\bibfnamefont {V.}~\bibnamefont {Garz\'o}},\ }\href {\doibase
  PhysRevE.49.367} {\bibfield  {journal} {\bibinfo  {journal} {Phys. Rev. A}\
  }\textbf {\bibinfo {volume} {49}},\ \unskip\ \bibinfo {pages} {367--375}
  (\bibinfo {year} {1994})}\BibitemShut {NoStop}%
\bibitem [{\citenamefont {Santos}(2009)}]{S09b}%
  \BibitemOpen
  \bibfield  {author} {\bibinfo {author} {\bibfnamefont {A.}~\bibnamefont
  {Santos}},\ }\href {\doibase 10.1007/s00161-009-0113-5} {\bibfield  {journal}
  {\bibinfo  {journal} {Cont. Mech. Thermodyn.}\ }\textbf {\bibinfo {volume}
  {21}},\ \unskip\ \bibinfo {pages} {361--387} (\bibinfo {year}
  {2009})}\BibitemShut {NoStop}%
\end{thebibliography}

%

\end{document}